\pdftrailerid{}
\documentclass[10pt,a4paper,twocolumn]{article}
\usepackage[margin=19mm,columnsep=6mm]{geometry}
\usepackage[T1]{fontenc}
\usepackage[utf8]{inputenc}
\usepackage{amsmath,amssymb}
\usepackage{booktabs,graphicx,microtype}
\usepackage[numbers,sort&compress]{natbib}
\usepackage{tikz}
\usetikzlibrary{arrows.meta,positioning}
\usepackage[hidelinks]{hyperref}
\usepackage{xurl}
\newcommand{\paperabstract}{%
Residual scalar quantization can produce highly dependent indices when successive stages reuse identical, unscaled grids. This case study examines that behavior in a frozen neural audio codec and develops an exact recoding procedure that reduces storage while preserving its decoder inputs. Analysis of the repeated-grid recurrence identifies predictable stage sequences; finite-precision checks motivate preserving every original index through explicit exception records. The model weights and reconstruction graph remain fixed. On the studied 62- and 512-file cohorts, a predictor, adaptive coding and explicit file selection reduce complete stored bytes. The final 512-file rate is 129.591~kbps including headers, 25.943\% below an original-index XZ control. Original recorded CPU runs establish bitwise latent and waveform preservation; independent readers verify exact index and metadata recovery in subsequent saved-stream experiments. Later coding choices use previously studied cohorts, and fresh encoding requires generating three alternatives. The engineering result is implementation-specific storage reduction with exact index recovery; the underlying lossy reconstruction quality remains unchanged.}

\title{Quantizer Redundancy and Exact Index Recoding in a Frozen Neural Audio Codec}
\author{Youssef Hemimy\\St.~Thomas University, Miami Gardens, Florida, USA\\\texttt{yhemimy@stu.edu}}
\date{13 October 2026}
\begin{document}
\twocolumn[\begin{@twocolumnfalse}
\maketitle
\begin{center}\small Preprint. Submitted to the \emph{Journal of the Audio Engineering Society}\newline as an Engineering Report on 13 October 2026. Not yet peer reviewed.\end{center}
\begin{abstract}\paperabstract\end{abstract}
\begin{center}\small\textbf{Keywords:} Entropy coding; finite scalar quantization; reconstruction fidelity; residual quantization.\end{center}
\vspace{1em}
\end{@twocolumnfalse}]
\section{Introduction}
Residual scalar quantization can produce highly dependent indices when successive stages reuse identical, unscaled grids. This case study examines that behavior in a frozen neural audio codec and develops an exact recoding procedure that reduces storage while preserving its decoder inputs. The practical question is how to identify the dependency, exploit it in a complete file format, and retain the behavior of the executed reconstruction graph.

The inspected codec uses separate low- and high-frequency signal paths, each with repeated scalar quantizer stages. Under its ideal-real recurrence, additional stages do not provide progressively finer resolution: odd-cardinality grids produce zero continuations, while even-cardinality grids produce alternating central contributions. For even-cardinality grids, the distortion law depends on stage parity and input. These properties identify a limitation of the implemented quantization rule and motivate a predictor for its stage indices; the analysis applies to these identical, unscaled grids.

Finite-precision execution adds a second engineering requirement. Ideal identities for accumulated grid values do not establish bitwise equality with the saved implementation. The recoder therefore checks predicted tuples against the actual indices and stores every exception in full. We evaluate preservation at three levels: integer indices, decoded scalar latents, and the neural decoder's waveform under a declared CPU execution. The weights, saved grids and reconstruction graph remain fixed.

The report combines the repeated-grid analysis, finite-precision characterization, and measured evaluation of an index-preserving format. A matched-zlib comparison isolates the representation change and its interaction with that compressor. The final format reaches 142.696 and 129.591~kbps on the studied 62- and 512-file cohorts, respectively, including headers: 26.296\% and 25.943\% below the development-selected original-index XZ control. Later coding choices use already studied cohorts, so these are descriptive case-study outcomes. The original CPU runs establish bitwise latent and waveform preservation; subsequent saved-stream checks establish exact restoration of their decoder inputs. The model's lossy reconstruction and historical quality limitations remain unchanged.

\section{Related Work and Scope}
SoundStream, EnCodec, and DAC establish encoder--quantizer--decoder designs for learned audio compression \cite{zeghidour2022soundstream,defossez2023encodec,kumar2023dac}. EnCodec also models discrete codes for entropy coding. Compression of dependent codec indices is therefore established practice. This case study uses fixed general-purpose compressor controls and classical adaptive probability models; it does not train an entropy model.

Finite scalar quantization replaces a learned vector codebook with low-dimensional scalar quantization and an implicit Cartesian-product code space \cite{mentzer2024fsq}. The inspected codec instead applies four scalar alphabet sizes across 128 channels per path, with symmetric, endpoint-inclusive uniform grids. Its nearest-grid rule differs from the shifted bounded-rounding construction in the original FSQ formulation. Parker et al. also describe post-training decomposition into scaled residual FSQ tokens~\cite{parker2025scaling}. The analysis below applies specifically to the present repeated, unscaled grids; it does not cover every FSQ or residual FSQ construction.

Our literature review includes primary records available through 6 October 2026. We acknowledge the earlier July 2026 robust residual FSQ revision on conditioning, the June 2026 preprint by Xu et al. on rate--distortion evaluation and recoding fixed audio indices, and the September 2026 preprint by Corlay et al. on conditional entropy coding with frozen transforms~\cite{zhu2026robustrfsqv4,xu2026benchmarking,corlay2026opportunistic}. These preprints are cited solely to acknowledge the temporal precedence of related ideas, not as evidence for this report's technical conclusions. The present report makes no priority claim for frozen-transform compression, post-training representation changes, or reconstruction-preserving index coding. Its technical results follow from the specified recurrence, finite arithmetic, and measured streams.

Integer entropy networks have previously addressed numerical disagreement across execution platforms~\cite{balle2019integer}. For the same temporal-priority purpose, we acknowledge the July 2026 MLVC report on deployment across heterogeneous hardware~\cite{parnamaa2026mlvc}. Our guarantees separately identify integer recovery, scalar latent bit equality, and waveform agreement under a declared execution; an ideal-real identity does not establish all three.

Objective waveform agreement also has a limited interpretation. SI-SDR measures a signal-domain relation \cite{leroux2019sdr}, and reference-based audio-quality predictions can depend on whether distortions preserve waveform correlation \cite{biberger2025quality}. The quality results retained here provide context for the frozen model. There is no listening study or rate-matched codec comparison in the new serialization experiments.

\section{Frozen Reconstruction System}
\subsection{Signal paths and quantizer configuration}
The system operates on mono audio at 44.1~kHz. A centered Hann STFT with length 2048 and hop 512 splits the spectrum at bin 185, approximately 3.984~kHz. Separate encoders use stride schedules $(8,8,4,2)$ and $(4,4,4,2)$, producing temporal hops of 512 and 128 samples. Both produce 128-channel latents bounded by a learned-scale hyperbolic tangent. Convolutional path decoders reconstruct bands that are recombined in the STFT domain. This reconstruction graph is unchanged throughout the experiments.

Each latent is divided into four groups of 32 scalar channels. The low-frequency path uses four stages with group alphabet sizes $(8,5,5,4)$; the high-frequency path uses six stages with sizes $(8,8,5,5)$. Every stage within a group uses the same saved grid. The selected checkpoint is at training step 595000. The experiments execute the preserved February source snapshot with strict state loading and frozen parameters. Historical comparison configurations are discussed separately in Sec.~\ref{sec:legacy}.

\begin{figure}[t]
\centering
\begin{tikzpicture}[
  >=Stealth,font=\footnotesize,
  box/.style={draw,align=center,minimum height=7mm,text width=6.8cm,inner sep=3pt},
  pathbox/.style={box,text width=3.1cm},
  arrow/.style={->,line width=.6pt}]
\node[box] (input) at (0,0) {Mono 44.1-kHz input\quad $\longrightarrow$\quad STFT band split};
\node[pathbox,anchor=north] (low) at (-1.85,-.95) {LF encoder: hop 512\\128 channels; 4 stages\\levels $(8,5,5,4)$};
\node[pathbox,anchor=north] (high) at (1.85,-.95) {HF encoder: hop 128\\128 channels; 6 stages\\levels $(8,8,5,5)$};
\draw[arrow] (input.south)--++(0,-.15)-|(low.north);
\draw[arrow] (input.south)--++(0,-.15)-|(high.north);
\node[box,anchor=north] (codes) at (0,-2.55) {Original stage-index tuples};
\draw[arrow] (low.south)--++(0,-.15)-|(codes.north);
\draw[arrow] (high.south)--++(0,-.15)-|(codes.north);
\node[box,fill=black!6,below=4mm of codes] (pack) {Predict tuples; compare every index\\Store normal records and complete exceptions};
\node[box,fill=black!6,below=4mm of pack] (stream) {Encode zlib, first- and second-order adaptive\\Select shortest; add 10-byte selector header};
\node[box,below=4mm of stream] (unpack) {Read mode; decode selected payload\\Restore original indices and metadata};
\node[box,below=4mm of unpack] (decode) {Saved-grid lookup and original stage-order sum\\Frozen LF/HF decoders and STFT recombination};
\node[box,below=4mm of decode] (output) {Preserved model reconstruction\\under unchanged decoding and execution};
\draw[arrow] (codes)--(pack);
\draw[arrow] (pack)--(stream);
\draw[arrow] (stream)--(unpack);
\draw[arrow] (unpack)--(decode);
\draw[arrow] (decode)--(output);
\end{tikzpicture}
\caption{Frozen reconstruction system with the final three-mode storage path. Gray boxes change index storage; saved grids and neural parameters remain unchanged. Independent readers verify restored indices and metadata for the final formats. Bitwise latent and waveform comparisons belong to the original CPU runs. No source audio or dataset-derived imagery is shown.}
\label{fig:architecture}
\end{figure}
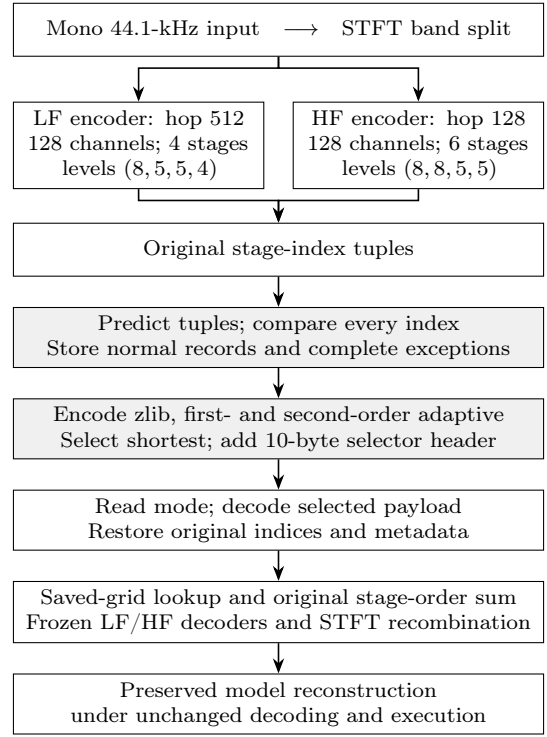

\subsection{The implemented scalar recurrence}
For a group of cardinality $q$, the ideal grid and its nearest-neighbor rule are
\begin{align}
 a_j&=-1+\frac{2j}{q-1},\quad j=0,\ldots,q-1,\\
 Q_q(r)&=a_{\arg\min_j|r-a_j|},
 \label{eq:grid}
\end{align}
with the smaller index selected at an exact tie. Starting from a bounded scalar $z\in[-1,1]$, the exact-real recurrence is
\begin{equation}
 r_0=z,\qquad v_\ell=Q_q(r_{\ell-1}),\qquad
 r_\ell=r_{\ell-1}-v_\ell.
 \label{eq:recurrence}
\end{equation}
The decoder receives $y_L=\sum_{\ell=1}^{L}v_\ell$. There is no reduction of grid scale with stage depth.

In the actual source, grid entries are float32 values and the forward quantizer uses the straight-through expression
\begin{equation}
 \widetilde v=r+\operatorname{stopgrad}(\widetilde a_j-r).
 \label{eq:ste}
\end{equation}
Code decoding instead looks up each saved grid entry directly and sums in stage order. Although Eq.~\eqref{eq:ste} equals the selected grid value in real arithmetic, finite-precision execution requires a separate forward-versus-code check.

\section{Repeated-Grid Analysis}
\subsection{Ideal arithmetic and stage parity}
A worked example illustrates the behavior before the general derivation. In exact arithmetic, take $z=1/5=0.2$ and reuse the four-level grid $\{-1,-1/3,+1/3,+1\}$ at every stage, with the nearest-grid rule in Eq.~\eqref{eq:grid}:
\begin{center}\small
\begin{tabular}{ccc}
\toprule
Stage $\ell$ & Contribution $v_\ell$ & Accumulated $y_\ell$\\
\midrule
1 & $+1/3$ & $1/3$\\
2 & $-1/3$ & $0$\\
3 & $+1/3$ & $1/3$\\
4 & $-1/3$ & $0$\\
\bottomrule
\end{tabular}
\end{center}
The accumulated output alternates between $1/3$ and zero. Here the second stage increases absolute scalar error from $2/15$ to $1/5$. This is an ideal-arithmetic example; the implemented float32 computation is examined separately below.

Let $h=1/(q-1)$, half the grid spacing. After the first nearest-grid operation, $r_1\in(-h,h]$ under the tie rule in Eq.~\eqref{eq:grid}.

For odd $q$, zero is a grid point and every residual in that interval quantizes to zero. All subsequent contributions vanish. For even $q$, the two central grid points are $-h$ and $+h$:
\begin{equation}
 Q_q(r)=\begin{cases}-h,&r\leq0,\\+h,&r>0,\end{cases}
 \qquad r\in(-h,h].
\end{equation}
Subtracting the selected central value changes the residual's sign interval. The next selected value has the opposite sign, and the residual returns to its preceding value after two stages. Consequently,
\begin{equation}
 y_L=\begin{cases}
 Q_q(z),&q\text{ odd or }L\text{ odd},\\
 Q_q(z)+Q_q(z-Q_q(z)),&q,L\text{ even}.
 \end{cases}
 \label{eq:parity}
\end{equation}
For these identical, unscaled grids, additional residual stages produce predictable continuations without progressively finer ideal resolution. The ideal final scalar has at most $q$ possible values on the closed input interval. For even $q,L$, these values are $-1-h+2jh$, $j=0,\ldots,q-1$; the lowest occurs only at $z=-1$ under the stated tie rule.

This result concerns ideal sums. For any tuple of indices, a weaker identity is
\begin{equation}
 y_L=-L+\frac{2S}{q-1},\qquad S=\sum_{\ell=1}^{L}j_\ell,
\end{equation}
so even arbitrary tuples admit a sum alphabet of at most $L(q-1)+1$ ideal values. Neither identity guarantees preservation of float32 accumulation or of individual stage indices. These dependencies motivate exact index recoding. The practical encoder compares predicted tuples with the emitted indices and preserves every mismatch in full, so finite-precision exceptions remain explicitly recoverable.

\subsection{Distortion depends on stage parity}
The same recurrence gives an exact distortion law. For even $q$, writing $e_\ell=z-y_\ell$ yields
\begin{align}
 |e_2|&=h-|e_1|,\\
 e_2^2-e_1^2&=h^2-2h|e_1|.
 \label{eq:distortion}
\end{align}
The second stage improves scalar squared error exactly when $|e_1|>h/2$, worsens it when $|e_1|<h/2$, and ties at equality. Odd depths repeat $e_1$ and even depths repeat $e_2$. For any input distribution, the mean squared error difference is $h^2-2h\mathbb{E}|e_1|$. Under a uniform input on $[-1,1]$, $\mathbb{E}|e_1|=h/2$ and both parities have mean squared error $h^2/3$. Learned encoder latents need not be uniform; these statements establish neither a preferred parity for the trained model nor a waveform-quality ordering. They also do not establish that this quantizer rule caused the historical full-band DAC gap. The supplement provides the proof and endpoint treatment.

\subsection{Nominal accounting versus serialized rate}
Allocating $\log_2q$ bits to each stage index gives
\begin{equation}
 R_{\mathrm{ideal}}=\sum_{p\in\{\mathrm{LF},\mathrm{HF}\}}
 \frac{44100}{h_p}L_p\,32\sum_{g=1}^{4}\log_2q_{p,g}.
 \label{eq:nominal}
\end{equation}
The declared configuration yields 810.415~kbps. Replacing logarithmic widths with whole-bit scalar fields gives an asymptotic payload of 915.075~kbps. Actual finite files also carry frame rounding and headers.

Using the ideal $q$-value final alphabet instead of all $L$ stages would give 143.929~kbps in logarithmic accounting, or 162.619~kbps in fixed scalar fields. These are analytical targets for a different representation. We do not use them as measured rates or assume that canonical grid substitution preserves the executed model.

\subsection{Complete finite-precision output classes}
\label{sec:classes}
For fixed saved grids and a specified arithmetic sequence, define $F(\mathbf j)$ as the float32 result of archived code decoding: the first grid lookup followed by successive additions in increasing stage order. Two tuples are equivalent when $F$ has the same bit pattern, including signed zero. A class identifier and a shared representative tuple then suffice to reproduce that scalar latent exactly. If the same latent tensor, shape, and metadata reach an unchanged deterministic neural decoder, its output is unchanged. This implication does not establish universal equality between the straight-through forward path and code decoding.

We exhaustively enumerate all $q^L$ tuples for each declared scalar configuration, at most 262144 per configuration. Table~\ref{tab:classes} gives the resulting class counts. Archived PyTorch decoding agrees bit-for-bit with independent sequential NumPy float32 addition for every tuple. Every class representative reproduces its members' decoded bits. These are complete results for the specified scalar code-decoding map, including tuples that may never be emitted by the encoder.

\begin{table}[t]
\centering\small
\caption{Exhaustive float32 code-decoder classes. $K$ counts all tuple outputs; $K_{\mathrm{n}}$ counts outputs of the current predictor's normal family. The latter contains $q$ tuples for odd $q$ and $2q$ for even $q$.}
\label{tab:classes}
\begin{tabular}{rrrrr}
\toprule
$q$ & $L$ & $q^L$ & $K$ & $K_{\mathrm{n}}$\\
\midrule
4 & 4 & 256 & 31 & 7\\
5 & 4 & 625 & 17 & 5\\
8 & 4 & 4096 & 105 & 15\\
5 & 6 & 15625 & 25 & 5\\
8 & 6 & 262144 & 211 & 15\\
\bottomrule
\end{tabular}
\end{table}

With a shared mapping, a standalone fixed-width class identifier requires $\lceil\log_2K\rceil$ bits. This does not specify an entropy rate or a complete file format. In particular, merging the normal families from 8 to 7 or from 16 to 15 outputs saves no whole fixed-width bit; the five-symbol families do not merge. Thus exhaustive classification strengthens the finite-arithmetic account without establishing a smaller practical bitstream. Unlike the following measured format, class-based substitution need not preserve the original stage indices. No audio result below uses it.

\section{Reversible Index Serialization}
\subsection{Predictor and escape rule}
For one scalar position, let $\mathbf j=(j_1,\ldots,j_L)$ be the indices actually emitted by the model and let $w=\lceil\log_2q\rceil$. For odd $q$, the predictor retains $j_1$ and predicts every later index as $\lfloor q/2\rfloor$. For even $q$, it retains $j_1$ and a branch bit indicating whether $j_2=q/2$; otherwise it predicts the lower central index $q/2-1$. Later predicted indices alternate between the two central indices.

The encoder compares the entire predicted tuple with $\mathbf j$. A matching tuple is represented by its first index and, when needed, the branch bit. Every mismatch has its position signaled and is stored as the complete original tuple. This includes noncentral second indices and any later deviation. The decoder reconstructs predicted tuples at unescaped positions and copies the complete tuples at escaped positions. Preservation therefore follows from exact integer comparison and reversible storage of exceptions; it does not require Eq.~\eqref{eq:parity} to hold in float32.

Let $J$ denote the original index arrays and $M$ the decoder-relevant metadata: sample count, sample rate, path frame counts, quantizer configuration, and checkpoint identity. For valid inputs supported by a format, the round-trip requirement is
\begin{equation}
 \operatorname{unpack}(\operatorname{pack}(J,M))=(J,M).
 \label{eq:roundtrip}
\end{equation}
Representation-specific fields such as payload lengths and integrity digests are outside $M$. Exactness holds even if every tuple escapes; sparse exceptions improve compression, not correctness. With the original lookup, accumulation, and neural decoding operations under the same execution conditions, restoring $(J,M)$ preserves the existing model reconstruction. Here lossless recoding refers to that reconstruction, not to the source audio already altered by the lossy model.

For a group with $N$ scalar positions and $E$ escapes, let $\delta_q=1$ for even $q$ and zero otherwise. The fixed-field payload is
\begin{equation}
 B=N+(N-E)(w+\delta_q)+ELw.
 \label{eq:payload}
\end{equation}
The first term is the escape bitmap. Escaped tuples do not also transmit the normal first-index and branch fields. If all tuples escape, the representation exceeds the original $NLw$ bits by $N$ flag bits. The fixed-field encoder reports that cost rather than selecting a different format.

\subsection{File format and compressor control}
The original and fixed-field predictive representations use a 120-byte header containing a format identifier and version, representation mode, sample rate, fixed quantizer configuration, original sample count, actual path frame counts, checkpoint digest, payload bit count, and integrity digest. Payload order is low-frequency groups followed by high-frequency groups. Within each group, channel--time positions are flattened with time varying fastest; stages remain in increasing order within a tuple. Integer fields are most-significant-bit first, with zero padding only at the final byte. The predictive payload stores the bitmap, normal records, and escaped tuples in that order for each group.

The companion supplement specifies every header field and its width, integrity scope, source revision, and checkpoint identity.

The baseline stores every original scalar index in $w$ bits. Both baseline and predictive files are also compressed independently with zlib level 6, using complete zlib wrappers and no shared or cross-file dictionary \cite{deutsch1996zlib,deutsch1996deflate}. This single preset is fixed across cohorts. All headers, checksums, padding, payload, and outer wrappers count toward measured bytes. The model is assumed shared; its weights are not retransmitted per clip.

Malformed lengths, unsupported configurations, out-of-alphabet indices, nonzero padding, and integrity failures are rejected. The format is a bounded research implementation. It is not presented as a production streaming protocol, and its digest is an integrity check rather than authentication.

\subsection{A constructive bound for sparse exceptions}
The fixed-field predictive format above uses one escape flag per scalar. As an analytical alternative, when $N$ is known, $E$ can be sent in $c_N=\lceil\log_2(N+1)\rceil$ bits and its position set ranked among $\binom{N}{E}$ possibilities. Keeping the same normal and full exception records gives
\begin{equation}
\begin{split}
 B_{\mathrm{enum}}={}&c_N+\left\lceil\log_2\binom{N}{E}\right\rceil\\
 &+(N-E)(w+\delta_q)+ELw.
\end{split}
\end{equation}
The position-rank field has zero bits for $E=0$ or $E=N$. This bound includes the exception count but excludes outer framing and final byte padding. A selector between bitmap and ranked-set encodings costs an additional bit and bounds their combined choice by the smaller payload plus that bit. It does not guarantee smaller zlib files, since zlib already compresses sparse flags. The initial zlib comparisons retain the original bitmap format; no result is assigned to combinatorial rank coding. The compression extension below instead evaluates practical sparse position lists.

\subsection{Adaptive coding and file selection}
Two predictive payloads use integer arithmetic coding~\cite{witten1987arithmetic}. Both use a 32-bit interval, initially uniform positive counts and count rescaling at a total of 16384. Exceptions are stored in full at delta-varint positions. For alphabet size $q$, the first-order model conditions each normal first index on the preceding coded normal first index within its channel. The second-order model instead uses context $a(q+1)+b$, where $a,b$ are the older and previous normal indices. Both states begin at sentinel $q$ at each channel boundary. Count vectors persist across channels within a stream and reset for each stream/file. Even-group branch coding is conditioned on the current first index in both models.

The final file selects the shortest complete payload among predictive+zlib6 ($Z$), first-order adaptive ($A_1$), and second-order adaptive ($A_2$), resolving ties in that order. An eight-byte magic, one version byte and one mode byte precede the unchanged payload. Consequently its length in bytes is
\begin{equation}
 \begin{split}
 L_3&=10+\min(Z,A_1,A_2),\\
 L_3&\leq L_2=10+\min(Z,A_1)\leq Z+10.
 \end{split}
 \label{eq:selector-bound}
\end{equation}
The bound includes signaling and holds separately for every file. It compares the final selector with its two-mode predecessor and the predictive+zlib payload, not with every compressor. All headers, lengths, exception records, termination, padding and integrity bytes are counted. The frozen neural model is assumed shared; no learned probability table or cross-file state is required. The supplement specifies each version and decoder validation rule.

\section{Validation Procedure}
\subsection{Cohorts and input handling}
Validation uses two cohorts: 62 files from the preserved evaluation and 512 additional files selected through inventory screening before neural inference. Both cohorts were subsequently reused during compression development; the final storage results therefore describe these studied collections.

The first cohort reuses the 62 reference files of the preserved electronic-music evaluation and contains 287.709~s of decoded audio. It establishes a new execution receipt for serialization; it does not recreate the undocumented historical software environment.

The additional screened cohort was selected from 183064 audio paths across 27 top-level collections. Normalized-path and conservative extensionless-basename exclusions against 14 available training and evaluation manifests left 38345 candidates. A deterministic round-robin procedure balanced top-level collections and preferred two-level pack keys absent from recorded training paths. It selected 512 byte-distinct files spanning 136 such keys. Screening checked prior evaluation hashes and size-matched accessible training files; 10580 training files were hashed for the candidates considered. Exact byte matches and selected duplicates were excluded.

This screening addresses overlap with the inspected training and evaluation inventories. Missing final-shard membership and augmentation-to-source mappings prevent a categorical unseen-content claim. The separate question of reuse during compression development is addressed below.

Inputs of at least 0.2~s were eligible. The additional screened cohort uses 398 complete files and one centered 10-s excerpt from each of 114 longer files. Its retained decoded duration is 2208.396~s. No selection decision depends on model output or measured rate. There are 46 mono and 466 stereo sources; stereo is downmixed by arithmetic mean, with no peak normalization or silence rejection.

SoundFile decodes float32 audio. The additional screened cohort includes 426 sources at 44.1~kHz and 86 at other rates between 42.928 and 192~kHz. Those 86 are converted to 44.1~kHz with SciPy polyphase resampling, a Kaiser window with parameter 5.0, and constant padding. Only the complete run with this preprocessing contributes to the results; the supplement records the earlier partial attempt and restart.

\subsection{Execution and equality checks}
Both original inference runs use the same checkpoint, archived model source, predictive format, and frozen CPU float32 execution. The environment includes PyTorch 2.9.1, four computation threads, and deterministic algorithms. Inference mode is enabled and gradients are disabled. Input hashes, source hashes, interpreter and package versions, configuration, checkpoint identity, and per-file stream hashes are recorded. All parameter and buffer hashes are compared before and after each complete run.

Table~\ref{tab:preservation-evidence} separates index restoration from the recorded neural-output comparisons. In the two original inference runs, forward latents and waveforms are compared with code-decoded outputs; decoding original codes is also compared with decoding unpacked codes. Float comparisons use exact bit patterns, not a numerical tolerance. Independent readers check saved-stream indices and metadata; reviewers inspect the recorded neural comparisons without independently repeating every inference.

\begin{table}[t]
\centering\small
\caption{Preservation evidence and its execution scope. The original inference runs cover both cohorts (62 and 512 files).}
\label{tab:preservation-evidence}
\begin{tabular}{@{}p{0.28\columnwidth}p{0.66\columnwidth}@{}}
\toprule
Preserved object & Evidence\\
\midrule
Indices and decoder metadata & Original versus unpacked values; independent saved-stream readers, including later formats.\\[3pt]
Latent bit patterns & Forward versus code-decoded, and original-code versus unpacked-code decoding, in the original CPU runs.\\[3pt]
Waveform bit patterns & The same two comparisons in the original CPU runs.\\
\bottomrule
\end{tabular}
\end{table}

Later formats are checked by restoring the same $(J,M)$ in Eq.~\eqref{eq:roundtrip}; they do not receive fresh neural inference runs. Their reconstruction-preservation claim depends on this exact restoration and unchanged decoding operations and execution conditions. Outer decompression must restore the complete inner stream byte-for-byte. After decoder padding is trimmed, rate is
\begin{equation}
 R_{\mathrm{measured}}=\frac{8\sum_i b_i}{\sum_i t_i},
\end{equation}
where $b_i$ counts all emitted bytes and $t_i$ is retained decoded duration. All natural outputs were long enough to retain the complete tested input or excerpt.

Selected scalar boundary cases additionally exercise endpoints, saved grid points, quantizer thresholds, and neighboring representable values for $q\in\{4,5,8\}$ at depths four and six. These are finite probes, not an exhaustive float32 proof. Eleven format tests cover normal and escaped tuples, invalid inputs, corrupted streams, and byte accounting. Independent review also checks stream accounting.

\subsection{Descriptive reconstruction metrics}
On the additional screened cohort, the archived implementations compute full-band SI-SDR and SI-SDR after an eighth-order Butterworth high-pass filter at 6~kHz applied forward and backward. Signals are trimmed from sample zero without lag search. These are waveform-agreement measures; the 6-kHz cutoff is not a brick-wall spectral boundary. Each clip has equal weight in descriptive metric means and medians. Input-energy diagnostics retain and flag silent references because the additive metric epsilon affects their scores.

\subsection{Development sequence and controls}
After cohort selection, the two original inference runs establish the index streams and compare original and predictive fixed-field representations under the same zlib level 6. Subsequent compression development reuses those saved indices. A later saved-index comparison predefines 14 candidates: four representations each under zlib6, Zstandard9 and XZ6, plus original-index and predictive adaptive coding. Full wrappers are counted; Zstandard includes content size and checksum, and XZ uses CRC64. The 62-file cohort selects one aggregate winner and one original-index control by total bytes, with source hashes and choices frozen before execution on 512 files. The 512-file execution applies choices frozen on 62 files, but both cohorts had already been studied. Later compression results are therefore descriptive outcomes on known data; this limitation is separate from incomplete evidence of training-content disjointness.

Two follow-ups use known earlier results: an explicit two-mode selector, then one fixed second-order context added as its third mode. The second-order candidate passes a positive-byte-saving gate for the three-mode selector on 62 files and is frozen before its 512-file execution. It changes only the first-index context. No context or preset sweep follows that result. Every new candidate and selected stream is checked against original indices and metadata, including independent decoder implementations. These passes reuse saved indices and do not rerun neural inference.

The supplement retains the complete candidate matrix and chronological results, including preserved failed or superseded attempts. A receipt-reader correction and a decoder-timing warmup correction each trigger renewed freezes and repeated development with byte-identical outputs before the affected 512-file pass. These repairs change neither coding choices nor reported file lengths.

\section{Results}
The final format reaches 129.591~kbps on the additional screened cohort, including all file headers and compression overhead. Rates below divide complete stored bits by total retained duration; the shared reconstruction model is excluded from per-file storage. The comparisons separate the representation effect, the final storage outcome, and the incremental benefit of adding the second-order candidate.

\subsection{Representation effect and preservation}
With identical zlib settings on both representations, predictive packing reduces bytes by 35.994\% on 62 files and 36.012\% on 512 (Table~\ref{tab:rates}). Every file becomes smaller in this comparison. Each of the additional screened cohort's 27 collection strata also shrinks, with approximately 34.9--41.5\% reduction in collection-level aggregate bytes. This matched-compressor comparison isolates the representation change and its interaction with that compressor; it does not claim a 36\% reduction against every compressor.

\begin{table}[t]
\centering\small
\caption{Measured rates, including all framing and compressor overhead. ``Original'' stores every stage index in fixed-width scalar fields.}
\label{tab:rates}
\begin{tabular}{lrr}
\toprule
Representation & 62 files & 512 files\\
 & \multicolumn{2}{c}{Rate (kbps)}\\
\midrule
Original & 915.523 & 915.751\\
Predictive & 245.587 & 245.666\\
Original + zlib & 240.692 & 218.670\\
Predictive + zlib & \textbf{154.057} & \textbf{139.923}\\
\bottomrule
\end{tabular}
\end{table}

The uncompressed predictive stream remains slightly larger than the zlib-compressed original stream in both cohorts. Complete compressed-versus-compressed controls therefore matter. On 512 files, predictive+zlib rates range from 1.002 to 189.195~kbps, with median 163.466~kbps; the aggregate 139.923~kbps includes silent or highly compressible material. Clip-level savings in this matched-zlib comparison range from 30.847\% to 53.420\%.

All indices are recovered exactly in the original two runs. Forward-versus-code and packed-versus-code latent and waveform comparisons are bitwise equal on all 574 files, and model-state hashes remain unchanged. There are two escapes among 15864832 scalar tuples on 62 files and ten among 121807104 on 512; all are preserved. Neither complete run has a failed item. Later formats recover these same indices and metadata, retaining the original stage lookup and summation order under the declared execution contract.

\subsection{Final storage outcome}
Table~\ref{tab:extension} places the final format alongside its controls and intermediate formats. The final selector stores 5131863 bytes on 62 files and 35773617 bytes on 512, at 142.696 and 129.591~kbps. Against the development-selected original-index XZ control, aggregate byte reductions are 26.296\% and 25.943\%. On the additional screened cohort, this 25.943\% saving preserves all decoder inputs, although one file remains 83 bytes larger than original+XZ.

Against predictive+zlib, aggregate byte reductions are 7.374\% and 7.384\%. On 512 files, 115 outputs exceed that unwrapped baseline by at most the selector's 10-byte header. Thus each aggregate saving and file-level comparison has a specific control. Percentages with different baselines cannot be added.

\begin{table}[t]
\centering\small
\caption{Complete compressed-file rates. XZ is the original-index control selected in the predefined 14-candidate comparison. Subsequent selectors and the second-order context are descriptive follow-ups on these known cohorts.}
\label{tab:extension}
\begin{tabular}{lrr}
\toprule
Representation & 62 files & 512 files\\
 & \multicolumn{2}{c}{Rate (kbps)}\\
\midrule
Original + XZ & 193.606 & 174.988\\
Predictive + zlib & 154.057 & 139.923\\
First-order adaptive & 148.610 & 137.184\\
Two-mode selector & 146.642 & 132.557\\
Second-order adaptive & 143.379 & 131.929\\
Three-mode selector & \textbf{142.696} & \textbf{129.591}\\
\bottomrule
\end{tabular}
\end{table}

Independent readers recover all original integer indices and decoder metadata from all 574 final candidate streams and all 574 selected outputs. These checks establish exact restoration of the decoder inputs. The bitwise neural-output checks above were performed in the original frozen CPU executions; the later saved-stream passes did not repeat neural inference.

\subsection{Where the additional bytes are saved}
The first-order adaptive candidate wins the 14-candidate development comparison, but its aggregate improvement hides regressions: on 512 files it makes 357 smaller and 155 larger than predictive+zlib, with a maximum increase of 34304 bytes. The two-mode selector avoids those payload increases at its explicit 10-byte framing cost. The complete candidate matrix and both cohorts' comparisons remain in the supplement.

First-index streams occupy 83.230\% of the first-order candidate's bytes on 62 files. Changing only their context to the preceding two coded normal indices reduces their stored bytes from 4448244 to 4260128 on 62, and from 31088663 to 29638245 on 512. Branch streams and complete exception records remain byte-identical across all 574 files. These first-index differences exactly account for the complete first-order versus second-order candidate differences.

The second-order candidate alone improves on first-order in 489 of the 512 cases, ties once and grows in 22. Its rate is 131.929~kbps before selection. Retaining both older modes therefore remains useful: the final 129.591~kbps result includes all selection overhead and does not assume that a longer context helps every file.

Adding the second-order candidate to the selector reduces aggregate bytes by 2.238\% on 512 files relative to the preceding two-mode selector: 378 files shrink, 134 tie and none grows. On 62 files, 55 shrink and seven tie. On 512 files, the chosen modes are 114 zlib, 20 first-order and 378 second-order. Both selectors include the same 10-byte header, so Eq.~\eqref{eq:selector-bound} guarantees this per-file non-expansion against the preceding selector. The comparisons with unwrapped predictive+zlib and original+XZ above retain their separate exceptions.

\subsection{Engineering cost and implementation limits}
A fresh final encoder generates three payloads from the same neural index arrays and stores the shortest. Saved indices support direct repacking. The decoder processes only the signaled payload before unchanged neural reconstruction. Adaptive probability models reset per file; neural weights remain fixed and shared. The format carries original length, path frame counts, configuration and checkpoint identity.

Table~\ref{tab:candidate-timing} reports warm serialization/encoding and decoding totals. Times exclude input unpacking, I/O, JIT compilation and neural inference. Selector experiments reuse older payloads, so complete three-mode encoding cost remains unmeasured.

\begin{table}[ht]
\centering\small
\caption{Warm candidate totals over 512 files. The first two rows share the original comparison run; second-order was measured separately.}
\label{tab:candidate-timing}
\begin{tabular}{lrr}
\toprule
Candidate & Encode (s) & Decode (s)\\
\midrule
Predictive + zlib & 7.949 & 6.524\\
First-order adaptive & 10.348 & 10.333\\
Second-order adaptive & 10.159 & 9.967\\
\bottomrule
\end{tabular}
\end{table}

Generating three alternatives and decoding one suggests offline storage and repacking as applications. These shared-machine timings do not establish production performance or latency. Process memory includes benchmark/JIT state and cannot isolate codec requirements. This bounded research codec implements integrity checking; streaming and random-access operation remain unevaluated.

\subsection{Finite-precision limits of canonical substitution}
Boundary probes produce six distinct float32 outputs for the four-level grid and fourteen for the eight-level grid, exceeding the ideal final alphabet sizes. The ideal recurrence consequently cannot be substituted blindly for the saved implementation. On the 62-file cohort, a separate diagnostic canonically reconstructs the ideal value associated with each actual emitted index sum. Every derived ordinal is in range, but all 62 waveforms change; the largest absolute sample difference is $1.654\times10^{-6}$. This approximation is excluded from the reported lossless recoding result. This diagnostic motivates the measured scheme's preservation of the original index tuple, including every exception, and the original lookup-and-sum order.

\subsection{Reconstruction quality on the additional screened cohort}
Table~\ref{tab:quality} characterizes the frozen reconstruction being preserved, measured against the preprocessed inputs. Mean full-band SI-SDR is 4.332~dB and its lower quartile is $-5.055$~dB, indicating substantial variation in waveform agreement across the additional screened cohort. Restoring the original decoder inputs preserves this existing reconstruction, and therefore these scores, under unchanged decoding and the declared execution conditions.

\begin{table}[t]
\centering\small
\caption{Additional screened cohort: objective metrics (512 files). Quartiles use linear interpolation. All scores are finite; silent references remain included.}
\label{tab:quality}
\begin{tabular}{lrrrr}
\toprule
Metric (dB) & Mean & Q25 & Median & Q75\\
\midrule
SI-SDR & 4.332 & $-5.055$ & 6.553 & 14.278\\
HF-SI-SDR-6k & 10.090 & 9.233 & 12.053 & 13.942\\
\bottomrule
\end{tabular}
\end{table}

Three inputs are exactly silent and four have high-pass RMS at or below $10^{-6}$. These overlapping diagnostics are retained rather than used for post hoc exclusions. Scores on zero-energy references depend on the metric's epsilon convention and have no direct fidelity interpretation. No significance or population-inference claim is attached to these descriptive summaries.

\subsection{Earlier objective evidence retained as context}
\label{sec:legacy}
The preserved historical 62-file records give mean full-band SI-SDR of 8.47~dB for the 595k model and 9.75~dB for DAC, with a paired mean difference of $-1.282$~dB. Their mean HF-SI-SDR-6k values are 11.64 and $-30.44$~dB. The historical full-band scores favor DAC at unmatched design points. The scalar quantizer analysis does not establish the cause of that gap. These records do not establish rate-matched superiority, and the new measured rate must not be presented as the rate of the historical DAC comparison.

Historical internal comparisons also mix training duration, architecture, capacity, quantization, and losses, with one saved run per configuration. On the 62-file cohort, the 140k model's onset-detection F1 comparisons against the three 100k variants have zero paired median differences and Holm-adjusted $p$ values of 0.955, 0.479, and 0.955. On the preserved 2144-file cohort, the 595k model has mean full-band and HF-SI-SDR-6k of 4.01 and 12.12~dB, while the comparison variants are at 100k steps. None of these comparisons isolates a causal component effect. The no-HF-loss variant also changes transient weighting, rather than removing a single spectral term alone.

The archived onset score uses detected onsets with 50-ms tolerance and a greedy matcher that can miss another available match after selecting an already-used estimate. It is not a direct attack-fidelity measure. Historical spectral-convergence and HF energy-fraction-ratio means are dominated by a few denominator-sensitive values. Those limitations, the full-band DAC advantage, and the transient null results remain part of the evidence. The historical aggregate tables and analyses are retained in the companion supplement; they are not newly executed quality comparisons.

\section{Discussion and Limitations}
The repeated-grid recurrence supplies a useful predictor, but the escape rule makes exact recovery independent of ideal arithmetic. The same standard compressor is applied to both representations, showing an additional approximately 36\% byte reduction attributable to the chosen representation and its interaction with that compressor in these experiments.

The theoretical results also constrain further simplification. Stage parity can improve or worsen ideal scalar distortion depending on the encoder distribution. Exhaustive enumeration characterizes exact output equivalence for the scalar code decoder, but does not characterize all reachable encoder states or universal forward/code agreement. In the normal predictor families, the only even-$q$ merge is between the two alternating central tuples that decode to positive zero. The remaining distinctions survive in float32, and the fixed-width token sizes do not decrease. The combinatorial sparse-exception bound is a constructive accounting result. The initial adaptive comparison combines practical sparse lists and probability modeling, so its extra saving cannot be assigned to that bound or to any one component. The subsequent second-order comparison changes only the first-index context; its byte decrease is fully accounted for in those streams.

The nominal 810.415-kbps count was therefore not a minimum storage requirement. Conversely, neither 143.929~kbps from ideal algebra nor the observed 129.591--154.057~kbps predictive compressed rates are universal codec operating points. The latter depend on actual symbol sequences, clip duration, framing, and compression. Other bit layouts, compressors, or entropy models could change the comparison. No preset sweep or optimality claim is made.

Preserving existing indices differs from changing the quantizer's resolution or training objective. Stage-dependent scaling could alter reconstruction and would require a separate study. Here, no neural parameter, saved grid, or decoder operation changes. The analysis should not be generalized to scaled residual FSQ, learned residual vector quantizers, or the original FSQ construction without examining their own rules.

Both cohorts are clustered convenience samples, and the additional sample is deliberately collection balanced. They do not represent population-random draws or independent training runs. Available inventories do not prove training-content disjointness. The observed exactness is tied to the recorded CPU float32 execution, and no other checkpoints, streaming conditions, or accelerators are evaluated. Runtime and memory measurements from the research scripts are not codec latency benchmarks.

The reconstruction model remains domain limited and mono. Stereo downmix and sample-rate conversion are preprocessing losses, independent of serialization. Objective fidelity varies substantially across additional inputs, and the historical full-band DAC advantage remains unresolved. No perceptual superiority, source-audio losslessness, or practical deployment readiness follows from the packing result.

\section{Conclusion}
This case study identifies avoidable index-storage cost in one frozen dual-path neural audio codec. Repeated, unscaled scalar grids create predictable stage continuations; complete exception records make exact index recovery independent of ideal-real identities.

The final format reaches 142.696 and 129.591~kbps on the studied 62- and 512-file cohorts, including all headers, reducing aggregate bytes by 26.296\% and 25.943\% against the selected original-index XZ control. Independent decoding verifies exact indices and metadata; original CPU executions establish latent and waveform preservation. A fixed temporal-context change accounts for additional savings in first-index streams, and explicit mode selection prevents per-file growth against the preceding selector with the same header.

For this frozen codec, quantizer redundancy can thus reduce storage while retaining every original decoder input. The gains describe cohorts reused during coding development and preserve the existing reconstruction under the declared execution conditions. Training-content lineage remains incomplete and complete encoding cost is unmeasured; broader codec generalization and deployment performance remain unestablished.

\section*{Data and Artifact Availability}
The manuscript and companion supplement provide mathematical descriptions, a source-independent schematic, file-format specifications, and aggregate results. Commercial source audio, identifying dataset records, per-file results, checkpoints, and implementation repositories are not publicly released. Execution and review receipts are retained internally. No release of these additional materials is promised.

\section*{Funding and Competing Interests}
No funding was received for this research. The author declares no competing interests.

\section*{Acknowledgments and AI Assistance}
OpenAI Codex (OpenAI; recorded client versions 0.160.1 and 0.162.0 and model identifier \texttt{gpt-6-astra}; used 6--8 October 2026) assisted with mathematical discussion, implementation of validation scripts, numerical analysis, literature checking, drafting, and preparation of the vector schematic. Further manuscript review and submission preparation on 13 October 2026 used Codex with recorded client version 0.162.1 and the same model identifier. These tools supported inspection and documentation of the frozen implementation and its recorded results. All empirical measurements reported here were produced by the recorded software executions. The author is responsible for the accuracy, originality, verification, and interpretation of all content.

\end{document}


\maketitle
\section*{Relationship to the Engineering Report}
This supplement retains the historical aggregate result tables and adverse, null, and denominator-sensitive findings from the earlier case-study draft. These are reaggregations of archived metric packages, not new evaluations of comparison codecs at the measured serialization rates. The new 62- and 512-file serialization experiments are reported in the main manuscript. They do not retroactively establish a historical evaluation environment or a matched-rate comparison with DAC.

Named checkpoints, reference audio, and evaluation manifests are available. February Git snapshots preserve the 595k evaluator and comparison evaluators alongside their result packages; exact historical execution receipts are absent, and the separate 140k producer invocation remains unlocated. The 595k and ablation-baseline 62-file packages agree on every shared numeric field and are treated as the same numerical evidence, not replications.

\section{New-Experiment Artifact Identity and Format}
Both original neural-inference validation runs use the preserved model-source revision
\nolinkurl{c768009f55ee00830979d162d40da8ea6a3c40f0}
and checkpoint SHA-256
\nolinkurl{3280beb5720bb8cad2c04b40dd87b113281a395ca21cf92010b837b2256a424e}.
These identifiers bind the new executions; they do not establish the inputs of an undocumented historical invocation. Model parameters and saved buffers have identical before/after hashes in both complete runs. The local source package contains no model weights or implementation repository.

\begin{table}[ht]
\centering\small
\caption{Version-1 inner header, in transmission order. Integers are unsigned big-endian. Offsets count bytes from zero; the complete header occupies 120 bytes.}
\begin{tabular}{rrl}
\toprule
Offset & Bytes & Field and required value\\
\midrule
0 & 8 & ASCII magic \texttt{EMCRCODE}\\
8 & 1 & Version: 1\\
9 & 1 & Mode: 0 original, 1 predictive\\
10 & 2 & Reserved flags: 0\\
12 & 4 & Sample rate: 44100\\
16 & 4 & LF stages, HF stages, groups, dimensions/group: 4, 6, 4, 32\\
20 & 4 & Two 16-bit hops: 512, 128\\
24 & 8 & LF then HF levels, one byte each: 8,5,5,4,8,8,5,5\\
32 & 8 & Original sample count before decoder padding\\
40 & 8 & Two 32-bit actual LF and HF frame counts\\
48 & 32 & Binary checkpoint SHA-256 digest\\
80 & 8 & Meaningful payload bit count, excluding final padding\\
88 & 32 & SHA-256 of bytes 0--87 concatenated with complete payload bytes\\
\bottomrule
\end{tabular}
\end{table}

The original payload traverses LF groups 0--3 then HF groups 0--3; each group traverses channel then time, with time varying fastest. At each scalar position it emits the complete stage tuple in increasing stage order, using $\lceil\log_2q\rceil$ bits per index. The predictive payload traverses the same groups. For each group, it emits all $N=32T$ escape bits, then all unescaped records in scalar order, then all escaped tuples in scalar order. An odd-$q$ normal record contains its first index. An even-$q$ normal record contains that index followed by one branch bit: zero selects second index $q/2-1$, one selects $q/2$. Later predicted indices alternate. Escapes store all original stage indices without duplicated normal fields. All integer fields are most-significant-bit first. Groups receive no byte alignment; zero bits pad only the last byte. The integrity digest covers those padding bits as well as the payload and prefix. The outer zlib stream compresses the entire inner header and payload without stripping any field.

\section{Additional Screened Cohort: Preprocessing Restart}
The 512-file selections were fixed before neural inference and retained across the preprocessing restart. The first attempt exposed excessive dense-filter cost in the original resampler for a 44053-Hz source. Its 37 completed receipts were retained separately; the same 512 selections were then evaluated afresh with the declared polyphase preprocessing. Only the complete restarted run contributes to the results. The restarted run uses SoundFile float32 decoding and, for the 86 sources outside 44.1~kHz, SciPy polyphase resampling with a Kaiser window of parameter 5.0 and constant padding. The partial attempt is preserved separately from the complete run and is not pooled into its reported results.

\section{Interpretation of Historical Metrics}
HF-SI-SDR-6k uses an eighth-order Butterworth high-pass filter with forward/backward filtering followed by zero-mean SI-SDR. Comparison signals are trimmed from sample zero without a lag search. ``Transient F1'' below denotes librosa onset-detection F1 with hop 512, backtracking, and a 50-ms tolerance; its greedy matcher can miss an alternative valid match after selecting an already-used estimate. Both-empty onset lists yield zero F1 in the retained implementation. The metric is not a direct measure of attack fidelity. HF energy preservation is a ratio of high-frequency energy fractions, rather than absolute retained HF energy.

Table intervals use deterministic 2000-replicate file bootstraps conditioned on the stored cohort. They are not population sampling intervals and do not capture training-run uncertainty. Paired signed-rank tests are aligned by stored cohort membership and Holm adjusted across each complete comparison table. No raw metric is recomputed or altered here.

\section{Preserved Historical Objective Results}
\subsection{Preserved 62-file cohort}
Table~\ref{tab:id} reports the three manuscript-safe primary metrics. Intervals are conditional file-bootstrap summaries; medians and IQRs expose skew.

\begin{table}[ht]
\centering\small
\caption{Aggregate objective results for the preserved 62-file cohort. Format: mean [95\% conditional file-bootstrap interval]; median [IQR].}
\label{tab:id}
\resizebox{\textwidth}{!}{%
\begin{tabular}{lccc}
\toprule
Configuration & SI-SDR (dB) & HF-SI-SDR-6k (dB) & Transient F1 \\
\midrule
MRHF@595k & 8.47 [6.37, 10.68]; 7.60 [2.42, 13.92] & 11.64 [10.52, 12.57]; 12.84 [11.06, 14.34] & 0.869 [0.807, 0.923]; 1.000 [0.826, 1.000] \\
DAC & 9.75 [7.37, 12.00]; 10.85 [5.11, 14.26] & $-30.44$ [$-33.61$, $-27.35$]; $-28.15$ [$-38.06$, $-21.72$] & 0.863 [0.795, 0.919]; 1.000 [0.859, 1.000] \\
MRHF@140k & 6.73 [4.56, 8.91]; 5.99 [1.81, 11.65] & 8.11 [6.97, 9.17]; 9.31 [7.46, 10.67] & 0.825 [0.757, 0.890]; 0.977 [0.756, 1.000] \\
SinglePath@100k & $-3.77$ [$-8.17$, 0.36]; 1.49 [$-12.59$, 6.21] & $-38.83$ [$-41.70$, $-35.81$]; $-37.81$ [$-46.38$, $-29.91$] & 0.820 [0.752, 0.883]; 0.971 [0.701, 1.000] \\
RVQ@100k & $-4.21$ [$-8.29$, $-0.23$]; 1.13 [$-10.35$, 5.05] & $-35.48$ [$-38.28$, $-32.90$]; $-35.34$ [$-41.32$, $-27.66$] & 0.803 [0.729, 0.870]; 0.969 [0.671, 1.000] \\
NoHFLoss@100k & 3.26 [$-0.04$, 6.50]; 6.04 [$-2.16$, 10.73] & $-38.16$ [$-41.10$, $-35.27$]; $-37.55$ [$-43.95$, $-29.69$] & 0.821 [0.748, 0.888]; 0.974 [0.761, 1.000] \\
\bottomrule
\end{tabular}}
\end{table}

The adverse full-band result is material: DAC's mean SI-SDR exceeds MRHF@595k by 1.282 dB. The paired MRHF-minus-DAC median is $-3.088$ dB (unadjusted $p=0.000599$; Holm-adjusted $p=0.00420$). In the same stored rows, mean HF-SI-SDR-6k is 11.64 dB for MRHF@595k and $-30.44$ dB for DAC, a mean paired difference of 42.08 dB. This is a design-point observation, not a rate-matched comparison.

The nearest-step internal comparison also contains null findings. For transient F1, MRHF@140k minus each 100k ablation has paired median zero. Unadjusted $p$ values are 0.477, 0.0958, and 0.596 for SinglePath, RVQ, and NoHFLoss; Holm-adjusted values are 0.955, 0.479, and 0.955. These results do not support a component effect on this transient metric. HF-SI-SDR-6k differences are large in the stored rows, but the 40k-step mismatch and one-run design prevent causal attribution.

\subsection{Preserved 2,144-file cohort}
A similar directional HF-SI-SDR-6k separation appears in the preserved seven-pack cohort (Table~\ref{tab:ood}). This is a cohort-specific observation, not independent replication. The MRHF@595k-versus-100k comparisons have a 495k-step mismatch.

\begin{table}[ht]
\centering\small
\caption{Aggregate objective results for the preserved 2,144-file cohort. Format follows Table~\ref{tab:id}.}
\label{tab:ood}
\resizebox{\textwidth}{!}{%
\begin{tabular}{lccc}
\toprule
Configuration & SI-SDR (dB) & HF-SI-SDR-6k (dB) & Transient F1 \\
\midrule
MRHF@595k & 4.01 [3.46, 4.54]; 4.64 [$-3.27$, 12.49] & 12.12 [11.92, 12.30]; 13.27 [11.17, 14.37] & 0.940 [0.935, 0.945]; 0.988 [0.938, 1.000] \\
SinglePath@100k & $-11.23$ [$-12.10$, $-10.26$]; $-3.31$ [$-26.15$, 5.22] & $-44.71$ [$-45.18$, $-44.23$]; $-43.74$ [$-51.37$, $-36.94$] & 0.901 [0.894, 0.908]; 0.968 [0.878, 1.000] \\
RVQ@100k & $-10.79$ [$-11.64$, $-9.93$]; $-3.12$ [$-25.33$, 4.96] & $-38.38$ [$-38.79$, $-37.98$]; $-37.15$ [$-43.35$, $-32.26$] & 0.901 [0.894, 0.907]; 0.967 [0.875, 1.000] \\
NoHFLoss@100k & $-3.52$ [$-4.30$, $-2.73$]; 2.01 [$-14.94$, 9.82] & $-39.00$ [$-39.45$, $-38.53$]; $-38.03$ [$-45.30$, $-31.88$] & 0.920 [0.914, 0.926]; 0.976 [0.906, 1.000] \\
\bottomrule
\end{tabular}}
\end{table}

No blanket significance threshold is used. For example, HF-energy preservation for MRHF@595k versus SinglePath@100k has unadjusted $p=4.94\times10^{-29}$, while transient F1 versus NoHFLoss@100k has unadjusted $p=5.95\times10^{-80}$ and Holm-adjusted $p=1.19\times10^{-79}$. The former belongs to a denominator-pathological metric family and is not interpreted as a stable mean effect.

\subsection{Denominator pathologies}
Arithmetic means for OOD spectral convergence, HF spectral convergence, and HF energy preservation are dominated by a few recurring members. Table~\ref{tab:pathology} contrasts means and medians for MRHF@595k. The largest 1\% of nonnegative values contribute approximately 99.997--99.9998\% of each sum. The same three hashed members recur among extremes across configurations and metrics. This pattern is consistent with near-zero reference-energy denominators. The reference audio is now available, but this audit did not recompute the denominators, so that mechanism remains unconfirmed. The affected arithmetic means are excluded from substantive comparisons.

\begin{table}[ht]
\centering\small
\caption{OOD denominator pathology for MRHF@595k; aggregate metric values only.}
\label{tab:pathology}
\begin{tabular}{lrrr}
\toprule
Metric & Mean & Median & Top-1\% sum share \\
\midrule
Spectral convergence & 68,139 & 0.147 & 99.9998\% \\
HF spectral convergence & 6,627 & 0.146 & 99.9974\% \\
HF energy preservation & 93,180 & 1.013 & 99.9989\% \\
\bottomrule
\end{tabular}
\end{table}

\section{Training and Comparison Limits}
The 27702-path selected source inventory reduces to exactly 24010 paths under the recorded minimum duration of 0.2~s; 3692 short files explain the difference. The broader parent inventory contains 98102 records. The combined and focused manifests contain 4621857 and 275527 rows, respectively, but their mixtures differ and the focused object is not an exact subset. Declared augmented row counts do not establish unique source recordings or actual training-shard consumption. Final shard membership and augmentation-source mappings remain incomplete, so neither archived ``ID/OOD'' label establishes training-content independence.

All comparison configurations have one stored training run and differ in rate, capacity, or optimization history. The no-HF-loss variant changes both spectral and transient weighting. The DAC folder label does not establish a measured bitstream rate. The 2144-file cohort has not received the new serialization validation, and no measured packing rate from another cohort is assigned to it.
\section{Distortion Law for the Ideal Repeated Grid}
Assume nearest-grid quantization with lower-index ties on the exact-real grid $a_j=-1+2j/(q-1)$, identical grids at each stage, and $z\in[-1,1]$. Set $h=1/(q-1)$ and $e_\ell=z-y_\ell$, where $y_\ell$ is the accumulated output. The first residual lies in $(-h,h]$.

If $q$ is odd, zero is a grid point and every subsequent contribution vanishes. If $q$ is even, the second-stage contribution is $-h$ for $e_1\leq0$ and $+h$ for $e_1>0$. Hence
\begin{equation}
 e_2=\begin{cases}e_1+h,&e_1\leq0,\\e_1-h,&e_1>0,\end{cases}
 \qquad |e_2|=h-|e_1|.
\end{equation}
The third contribution has the opposite sign from the second, returning the residual to $e_1$. Thus all odd depths share $e_1$ and all even depths share $e_2$. Squaring the absolute-error relation gives
\begin{equation}
 e_2^2-e_1^2=h^2-2h|e_1|.
\end{equation}
This includes $e_1=0$, where the lower-index tie gives $e_2=h$, and $e_1=h$, where $e_2=0$. Both parities have worst-case absolute scalar error $h$. For any input distribution, including atoms, the expected squared-error difference is $h^2-2h\mathbb{E}|e_1|$. Additional depth can therefore improve or worsen expected error depending on that distribution.

For a uniform input on $[-1,1]$, the $q-2$ interior first-stage cells have width $2h$ and the two endpoint cells each have width $h$. Integration gives
\begin{align}
 \mathbb{E}e_1^2&=\frac12\left[(q-2)\frac{2h^3}{3}+2\frac{h^3}{3}\right]=\frac{h^2}{3},\\
 \mathbb{E}|e_1|&=\frac12\left[(q-2)h^2+h^2\right]=\frac h2.
\end{align}
Consequently every depth has mean squared error $h^2/3$ under this illustrative distribution. This does not assume that the trained encoder produces uniform latents. For example, an even-$q$ input concentrated at zero has $e_1=h$ and $e_2=0$, favoring even depth. Conversely, an input at an original grid point has $e_1=0$ and $|e_2|=h$, favoring odd depth.

At even depth and even $q$, inputs between adjacent original grid points reconstruct their interval midpoint, with the lower midpoint selected at a shared boundary. The isolated endpoint $z=-1$ instead reconstructs $-1-h$. It has zero mass in the continuous uniform example, but cannot be omitted from a closed-domain or floating-point endpoint claim. None of these scalar latent statements establishes an ordering of waveform distortion after neural decoding.

\section{Exhaustive Finite Arithmetic Certificate}
The scalar enumeration uses the same archived quantizer source and saved-grid bit patterns identified above, with Python 3.13.9, PyTorch 2.9.1 and NumPy 2.3.5 on an ARM64 CPU. It enables deterministic algorithms and uses one Torch thread. It does not run an audio encoder, neural decoder, or straight-through forward quantizer. Saved grids are reconstructed from the earlier hashed checkpoint probe receipt; that receipt confirms bit identity across the relevant groups and stages.

The following hexadecimal words specify the saved IEEE 754 binary32 grids completely, in increasing index order. They describe the sign, exponent, and fraction bits of each value, independently of memory byte order. They must not be replaced by separately rounded ideal rational values. Decode by starting with the first indexed word and rounding each subsequent addition to binary32 with round-to-nearest, ties-to-even; do not reassociate additions.
\begin{center}\small
\begin{tabular}{rl}
\toprule
$q$ & Ordered hexadecimal grid words\\
\midrule
4 & \texttt{bf800000 beaaaaaa 3eaaaaaa 3f800000}\\
5 & \texttt{bf800000 bf000000 00000000 3f000000 3f800000}\\
8 & \texttt{bf800000 bf36db6e bedb6db6 be124922}\\
  & \texttt{3e124922 3edb6db6 3f36db6e 3f800000}\\
\bottomrule
\end{tabular}
\end{center}

For each declared $(q,L)$ pair, tuples are enumerated in lexicographic order, represented by their base-$q$ ranks. The archived scalar code decoder looks up the first saved grid value and then adds each later lookup in stage order, in float32. An independent NumPy transcription uses explicit sequential float32 additions, with no reduction reassociation. Every output bit pattern agrees between these two calculations. The five tuple spaces contain respectively 256, 625, 4096, 15625 and 262144 tuples, totaling 282746.

Define $\mathbf j\sim\mathbf k$ if their code-decoded scalar outputs have identical 32-bit patterns. Classes are ordered by the unsigned integer interpretation of those patterns. The representative of each class is its lexicographically first tuple. Exhaustive checks establish that every tuple's representative decodes to its original bits, every representative maps back to its class, and class membership counts sum to $q^L$. Signed zeros, if present, are treated as different patterns. Input/source hashes, class maps, membership counts, representative ranks, script identity and runtime receipts are retained internally.

\begin{table}[ht]
\centering\small
\caption{Finite arithmetic counts and standalone field widths. The ideal sum column permits arbitrary tuples; it is different from the smaller alphabet of the bounded exact-real recurrence.}
\begin{tabular}{rrrrrr}
\toprule
$q$ & $L$ & Ideal unrestricted sums & Float32 classes & Original bits & Class bits\\
\midrule
4 & 4 & 13 & 31 & 8 & 5\\
5 & 4 & 17 & 17 & 12 & 5\\
8 & 4 & 29 & 105 & 12 & 7\\
5 & 6 & 25 & 25 & 18 & 5\\
8 & 6 & 43 & 211 & 18 & 8\\
\bottomrule
\end{tabular}
\end{table}

The arbitrary-tuple alphabet includes tuples that the residual encoder may never emit. It therefore does not establish the smallest reachable alphabet. For the odd-$q$ normal predictor family, all later indices are central zeros. For even $q$, the first index and either central second index determine the alternating continuation. Their float32 class counts are 5 for $q=5$, 7 for $q=4$, and 15 for $q=8$ at the declared depths. The reductions from 8 to 7 and 16 to 15 normal tuples save no whole fixed-width bit. A smaller class alphabet can affect another coding scheme, but does not imply a smaller zlib-compressed file.

A shared class map can reproduce an exact latent by selecting a representative tuple. It need not recover the original indices or intermediate-stage sums. The deterministic neural-output implication requires unchanged latent bits, shape, metadata and subsequent decoder execution. No universal equivalence with the encoder's straight-through expression or another hardware/runtime is asserted. The audio experiments continue to use the original index-preserving format.

\section{Constructive Sparse Exception Bound}
For known $N$, store the exception count $E$ in $c_N=\lceil\log_2(N+1)\rceil$ bits. There are $\binom{N}{E}$ distinct exception position sets. An agreed ordering assigns each set a rank requiring $c_P=\lceil\log_2\binom{N}{E}\rceil$ bits, with zero bits if only one set exists. Invalid ranks are rejected. This construction assumes no statistical independence.

With $b=w+\delta_q$, the complete inner record payload is
\begin{equation}
 B_{\mathrm{enum}}=c_N+c_P+(N-E)b+ELw.
\end{equation}
Compared with the present bitmap payload, the exact bit saving before framing and byte padding is $N-c_N-c_P$. The usual counting inequality $\log_2\binom{N}{E}\leq N H_2(E/N)$ gives an upper bound in terms of binary entropy, with $0\log_2 0=0$. It is not a measured entropy rate of the codec.

Choosing the smaller bitmap or ranked-set payload needs a selector unless a paid format field already identifies the choice. One explicit selector bit gives $1+\min(B_{\mathrm{bitmap}},B_{\mathrm{enum}})$. Likewise, allowing the original fixed-field tuples as a fallback gives $1+\min(NLw,B_{\mathrm{alternative}})$ before common framing and padding. The selector prevents claiming a strict no-expansion guarantee without accounting for its cost. None of these inner-stream bounds guarantees improved compression after zlib; an actual format comparison must count every transmitted byte. The initial audio-run rates use the original bitmap format. The subsequent saved-stream extension uses practical position lists; combinatorial rank coding and the selector alternatives above remain analytical constructions.

\section{Bounded Saved-Stream Compression Extension}
The extension consumes the same saved original and predictive integer streams; it does not read audio, load weights, or execute a neural encoder or decoder. Development uses all 62 existing files and confirmation all 512 existing files. These cohorts were previously examined. The candidate family, presets, aggregate-byte selection rule, and tie-breaks were specified before development. There is no parameter or context sweep beyond the declared 14 candidates.

For every file, the original v1 stream is decoded to integer arrays. Each representation receives those same arrays. The legacy original and predictive encoders reproduce their recorded streams byte-for-byte. Complete compressed files are written and reread; standard decompression must recover the complete inner bytes, and every candidate must recover all integer arrays, actual frame counts, sample count, sample rate, fixed model configuration and checkpoint identity. The shared model is excluded from per-file costs consistently.

\subsection{Representations and standard compressors}
The four representations are original v1, predictive v1, separated predictive with bitmap escapes, and separated predictive with sparse absolute-position lists. Separated variants retain the 120-byte header layout, replacing its magic with \texttt{EMCRSEPR}, version 1 and mode 0 (bitmap) or 1 (sparse). Prefix/payload integrity and fixed configuration checks are unchanged. For each group the bitmap mode stores $N$ flags; the sparse mode stores $E$ in $\lceil\log_2(N+1)\rceil$ bits and $E$ strictly increasing positions in $\lceil\log_2N\rceil$ bits each. Both then store normal first indices, normal even-group branch bits, and full exception tuples as three separate regions. Whole-tuple comparison determines exceptions; group boundaries have no extra alignment and only the final byte is padded. Sparse lists are not combinatorial rank coding.

Each representation is compressed separately per file with zlib level 6, Zstandard level 9 (single thread, content size and checksum enabled), or LZMA/XZ preset 6 (CRC64). The versions are zlib runtime 1.3.1 and Zstandard library 1.5.7 through Python zstandard 0.24.0; XZ uses Python 3.13.9's standard-library LZMA interface. Complete wrappers are included. No dictionary or state is shared across files.

\subsection{Adaptive arithmetic format}
Both adaptive modes use a 120-byte header with the same configuration and identity fields, magic \texttt{EMCADAPT}, version 1, and variant 0 (original) or 1 (predictive). The payload-bit field is eight times the complete payload byte length; SHA-256 covers the prefix and payload. The implementation bounds accepted outputs to 16 million indices per file. No learned entropy model or fitted probability table is required by this arithmetic coder; the frozen neural reconstruction model remains shared.

The arithmetic coder maintains an inclusive 32-bit interval $[\ell,u]$, initially $[0,2^{32}-1]$. For total count $t$, cumulative counts $a,b$ bounding the symbol, and width $v=u-\ell+1$, simultaneous updates are
\begin{equation}
 u'=\ell+\left\lfloor\frac{vb}{t}\right\rfloor-1,
 \qquad \ell'=\ell+\left\lfloor\frac{va}{t}\right\rfloor.
\end{equation}
Products use 64-bit integers. Standard half-interval and middle-half renormalization emits bits and defers underflow complements. Each alphabet count starts at one. After a symbol the selected count increases; when a context total reaches 16384, every count $c$ becomes $(c+1)\mathbin{//}2$. Models reset for each stream and file. A nonempty stream terminates by adding one pending bit, emitting zero if $\ell<2^{30}$ and one otherwise, then all pending complements, 32 explicit zero lookahead bits, and final zero byte padding. Empty streams have zero bytes. Each arithmetic stream carries an eight-byte big-endian length.

First-stage indices use the previous coded first index within the channel as context, with sentinel $q$ at each channel boundary. Counts persist across channels within that stream, but the previous-symbol state resets. Original mode codes each later stage with its own unconditional model. Thus its context model is a declared control, not an optimized original-index entropy model. Original mode has 40 streams, one per path/group/stage.

Predictive mode compares complete tuples with the same legacy predictor. For each group it writes an eight-byte exception count, an eight-byte exception-position byte length, positive position deltas encoded as minimum-length unsigned base-128 varints, full exception tuples in fixed scalar fields with zero padding, the normal-first arithmetic stream and, for even groups, the normal-branch arithmetic stream. Position deltas start from previous position $-1$. Normal-first context uses the previous normal first index in the same original channel; exceptional positions do not update that state. Branch context is the current decoded first index. There are eight normal-first streams and four branch streams. All normal and exceptional positions are reconstructed in their original order.

NumPy 2.3.5 coordinates arrays; Numba 0.62.1 compiles integer coding loops without disk caches. Synthetic checks include normal/mixed/all exceptions, channel-boundary cases, repeated rescaling, noncontiguous inputs, metadata, byte accounting and corruption. Independent review uses separate full-stream parsers and a separately transcribed arithmetic decoder without importing producer modules. The complete 2048 confirmation streams recover their source arrays and metadata exactly; all ten exceptional tuples remain present. Waveform preservation follows from restoring the established decoder inputs under its unchanged execution contract; no additional waveform inference is claimed.

\subsection{Selection and complete development matrix}
Table~\ref{tab:compression-development} reports every predefined development candidate. The global winner minimizes total file bytes; exact ties would use aggregate decoding time, then lexical candidate identifier. The original-index control is selected separately by the same rule. Adaptive predictive wins; original+XZ is the selected control. Neither per-file oracle selection nor a fallback selector is applied in that experiment.

\begin{table}[ht]
\centering\small
\caption{Complete development results (62 files, 287.709 s). Every candidate recovers all indices and metadata exactly. Rates include all transmitted bytes.}
\label{tab:compression-development}
\begin{tabular}{lrr}
\toprule
Candidate & Bytes & Rate (kbps)\\
\midrule
Adaptive predictive & 5344543 & 148.610\\
Predictive + XZ & 5404552 & 150.278\\
Predictive + zlib & 5540433 & 154.057\\
Separated sparse list + XZ & 5594104 & 155.549\\
Separated bitmap + XZ & 5601264 & 155.748\\
Predictive + Zstandard & 5643730 & 156.929\\
Separated sparse list + zlib & 5714815 & 158.906\\
Separated bitmap + zlib & 5731546 & 159.371\\
Separated sparse list + Zstandard & 5799574 & 161.262\\
Separated bitmap + Zstandard & 5812950 & 161.634\\
Original + XZ & 6962772 & 193.606\\
Original + Zstandard & 8399487 & 233.555\\
Original + zlib & 8656130 & 240.692\\
Adaptive original & 9051576 & 251.687\\
\bottomrule
\end{tabular}
\end{table}

The original confirmation attempt reads the first saved pair but stops before any candidate encoding because that cohort's hash receipt is nested differently. The failure, initial development outputs and freeze are preserved. Only the receipt adapter is repaired. All 868 development files (62 times 14 candidates) are then reproduced byte-for-byte and a new source/configuration freeze precedes confirmation. No candidate or compressor changes, and no confirmation compression outcome informs the new freeze.

On confirmation, adaptive predictive stores 37869472 bytes, predictive+zlib 38625631, original+XZ 48305372, and original+zlib 60363622. Retained duration is 2208.396 s. The adaptive gain over the incumbent is 756159 bytes (1.958\%); 357 files shrink and 155 grow. The largest increase is 34304 bytes. Against original+XZ, 472 shrink, one ties and 39 grow. This rules out a universal file-size improvement claim.

Warm adaptive encoding/decoding takes 10.348/10.333 s over the confirmation cohort, compared with 7.949/6.524 s for predictive+zlib and 83.733/14.009 s for original+XZ. Encoding includes index serialization and compression, but excludes legacy input unpacking, disk writes, initial JIT compilation and neural inference. These are single shared-machine executions; process peak memory includes the benchmark and JIT and is not an isolated codec-memory measurement. The modest byte reduction does not establish that adaptive coding should replace the incumbent for every use case.

\section{Two-format selector follow-up}
After the fixed-candidate results are known, a separately authorized follow-up fixes the rule $10+\min(Z,A)$, where $Z$ and $A$ are the complete predictive+zlib6 and adaptive-predictive byte lengths. Ties choose zlib. Source and protocol hashes are frozen before this measurement on either cohort. This ordering does not make previously studied cohorts an untouched holdout.

The wrapper is ASCII \texttt{EMCSELCT} (eight bytes), version 1 (one byte), mode (one byte: 0 for zlib, 1 for adaptive), followed by the unchanged selected payload to the file boundary. It adds no other digest: the selected inner stream retains its original integrity check and metadata. The decoder rejects unsupported headers, modes, mismatched inner formats, truncation and trailing inner bytes. It requires neither the unselected payload nor the original indices. Shared neural reconstruction weights remain required for audio decoding.

Development stores 5273781 bytes including 620 header bytes, with 51 adaptive and 11 zlib choices. The 512-file cohort stores 36592418 bytes including 5120 header bytes, with 357 adaptive and 155 zlib choices. On that cohort, equally wrapped zlib costs 38630751 bytes and equally wrapped adaptive costs 37874592 bytes. The selector reduces those totals by 5.276\% and 3.385\%, respectively, and is never larger per file than either equally wrapped alternative. Relative to unwrapped zlib, 357 files shrink and 155 grow by exactly 10 bytes. Relative to unwrapped adaptive, 155 shrink and 357 grow by exactly 10 bytes. Relative to original+XZ, 511 shrink and one grows by 83 bytes.

Both modes, from-indices encoding, ties, the length bound, and malformed inputs are checked synthetically. Every saved output is reread and decoded against the original baseline indices and metadata. A separate wrapper parser with independently implemented inner decoders checks all 574 outputs. Saved payloads are reused without recompression or neural inference. Consequently this execution supplies no fresh complete encoding-time measurement. A real encoder must generate both alternatives; the sum of their earlier 512-file encoding times is 18.297~s, which is only a historical cost estimate excluding selection and wrapping. No speed improvement or production latency is claimed.

\section{Second-order context follow-up}
The next experiment introduces exactly one candidate, retaining the prior adaptive format except for magic \texttt{EMCAORD2}, predictive-only operation, and the first-index probability context. At a channel boundary, older and previous normal-symbol states are $(q,q)$. The context is $a(q+1)+b$, indexing $(q+1)^2$ initially uniform count rows. After coding symbol $s$, states become $(b,s)$. Count vectors persist across channels within one stream, but the two symbol states reset; exceptions remain omitted. No backoff, mixture, fitted table or parameter sweep is used. Branch coding, interval arithmetic, rescaling, exceptions, framing and integrity remain unchanged.

Selector version 2 keeps the same eight-byte magic \texttt{EMCSELCT}, version byte and mode byte. Modes 0, 1 and 2 identify zlib, first-order and second-order payloads. The lowest byte length wins, breaking ties by lower mode. The new decoder also accepts version-1 modes 0 and 1; version-1 mode 2 is rejected. Full selected payloads are retained. Consequently every output is no larger than the previous two-mode selector, with all header bytes included.

The first 62-file execution used an all-exception warmup fixture, which could leave decoder JIT compilation inside the first measured decode time. Independent review identified this before the new candidate ran on 512 files. Only the fixture was corrected to contain normal predictor-consistent symbols in every group. The original harness, freeze and outputs were preserved; a new freeze preceded a repeated 62-file run. All 124 emitted candidate/selected development files reproduced byte-for-byte. The repeated development result passed the fixed positive-byte-saving gate, and its receipts and unchanged source were frozen before the 512-file run.

The new candidate stores 5156427 bytes on 62 files and 36419054 bytes on 512; the three-mode selector stores 5131863 and 35773617 bytes, including 620 and 5120 header bytes. On 62, the choices are seven zlib, zero first-order and 55 second-order. On 512 they are 114, 20 and 378. Relative to the previous selector, 55 development files shrink and seven tie; 378 confirmation files shrink and 134 tie. None grows. Relative to old unwrapped zlib on 512, 397 shrink and 115 grow by at most 10 bytes; the latter include one selected second-order payload whose saving is less than the header. Original+XZ remains smaller for one file by 83 bytes.

Accounting isolates the changed coding component: first-index streams fall from 4448244 to 4260128 bytes on 62, and from 31088663 to 29638245 bytes on 512. All branch and complete exception records are byte-identical; their encodings and lengths do not explain the gains. These differences exactly equal the total first-order versus second-order candidate differences. Independent verification checks all 574 candidates and all 574 selected outputs against the original indices and metadata. No audio or model execution occurs. The 10.159/9.967~s warm encoding/decoding measurements cover only the new 512-file candidate; old payloads are reused. They do not measure full three-mode encoding or establish production performance.